\documentstyle[12pt]{article}
\begin{document}
\title{Testing ETC  Generation of the Top Quark Mass}
\renewcommand{\baselinestretch}{1.0}
\author{Lisa Randall\thanks{This work is supported

in part by funds provided by the U.S.

Department of Energy (DOE)

under contract \#DE-AC02-76ER03069 and in

part by the Texas National Research Laboratory Commission
under grant \#RGFY92C6.\hfill\break
National Science

Foundation Young Investigator Award.\hfill\break
Alred P.~Sloan Foundation Research Fellowship.\hfill\break
Department of Energy Outstanding Junior

Investigator

Award.\hfill\break
CTP\#2111\hfill September 1992}\\
Massachusetts Institute of Technology\\
Cambridge, MA 02139\\
}
\date{}
\maketitle
\vskip-5in
May 1992 \hfill  MIT-CTP\#2111
\vskip5in
\renewcommand{\baselinestretch}{1.2}
\abstract{
We consider constraints on models in which a top quark
mass is generated  through

unenhanced extended
technicolor interactions. The deviation in the $\rho$
parameter

from  unity and $B$--$\overline{B}$ mixing could be large,
but given the
uncertainties in strong dynamics and variations in the parameters of
models, no conclusive statement can be given. We conclude that the

low technicolor scale which
is required to generate the top quark
mass is not ruled out.
}
\thispagestyle{empty}
\newpage

\newcommand{\hc}{{\rm h.c.}}
\newcommand{\gev}{{\rm GeV}}
\newcommand{\mev}{{\rm MeV}}
\newcommand{\tr}{\mathop{\rm tr}\nolimits}
\newcommand{\uone}{\mbox{U(1)}}
\newcommand{\su}[1]{\mbox{SU(#1)}}
\newcommand{\cf}{{\it c.f.}}
\newcommand{\half}{{\textstyle{1\over2}}}

\newcommand{\third}{{\textstyle{1\over3}}}

\newcommand{\fourth}{{\textstyle{1\over4}}}

\newcommand{\bra}[1]{\left\langle #1\right|}
\newcommand{\ket}[1]{\left| #1\right\rangle}
\newcommand{\vev}[1]{\left\langle #1\right\rangle}
\newcommand{\hatk}{\hat{k}}
\newcommand{\hatq}{\hat{q}}
\newcommand{\hatl}{\hat{l}}
\newcommand{\imn}{{I_{\mu \nu}}}
\newcommand{\beq}{\begin{equation}}
\newcommand{\eeq}{\end{equation}}
\def\barr{\begin{eqnarray}}
\def\ear{\end{eqnarray}}
\newcommand{\ba}{\begin{array}}
\newcommand{\ea}{\end{array}}
\section{Introduction}

It is well known that extended technicolor \cite{ETC} is problematic.

Although
technicolor \cite{TC} works well as a mechanism for spontaneously

breaking the
electroweak symmetry, no such aesthetic appeal can be claimed for
the flavor dynamics that are required in technicolor scenarios. The
fundamental problem is to break the flavor symmetries among quarks
and leptons without simultaneously generating too large flavor

changing
neutral currents \cite{TC,wellknown}.

For this reason, alternatives to the simplest
extended technicolor theories have been proposed; walking technicolor
\cite{WTC}
is a  well studied example. However, even in theories where
the technicolor dynamics is assumed to be very different

from that of  QCD,
the top quark mass might not be significantly enhanced by
walking because of the low extended technicolor scale

\cite{lizsek}. For the purposes
of this analyis, we make the same assumption as in reference

\cite{lizsek},
that the top quark gets its mass from extended technicolor with no
enhancement, either from fine tuning \cite{fine} or from walking

\cite{WTC}. Our results should apply to most other models in which
the flavor sector distinguishes the top quark mass.
In ref. \cite{lizsek}, it was shown that the coupling of $b$ quarks
to the $Z$

can get a measurable correction from extended technicolor

interactions. In this paper, we consider  other constraints which
must be considered if one assumes an extended technicolor scale
which is sufficiently low to give the top quark a mass. We first
show that $\Delta \rho$, the deviation in the $\rho$ parameter from
unity, is large. This will probably prove
a stronger constraint than the precision measurement of the
$b$ quark vertex, discussed in ref. \cite{lizsek},
 although this could prove a useful confirmation. We then
derive the consequences of flavor
dynamics at this scale for flavor changing neutral currents. We show
that when there is a measurable change in the $Z b \overline{b}$

vertex,
it will generally be accompanied by large flavor changing neutral

current
effects. These are at the border of acceptibility.

Our work differs from previous analyses of extended technicolor
for the following reasons. First, we consider only the
effect from the interactions necessary to generate the top quark
mass.
It is well known that simple ETC does not work for the

light   generations.
Second, in addition to flavor changing effects which are directly
induced through light quark couplings of the ETC gauge bosons,
we also consider flavor changing $Z$ couplings which arise from
a light fermion--technifermion vertex.  Third, we put all the
constraints together to determine whether an ETC scale as low
as would be necessary for the top quark is allowed.

Our conclusion is perhaps surprising. We find that although
there are large corrections to the $Z$ vertex, probably most
of the flavor changing effects which are generated are comparable
to those of the standard model or else will not be measured.
The strongest constraints turn out to be the constraint from
the $\rho$ parameter, previously considered in ref. \cite{appel}
and from $B-\overline{B}$ mixing, which is similar to the kaon
mixing constraint on ETC for the light generations.  However,
because of strong interaction uncertainties, one would need to
disagree with experiment by an order of magnitude before one could
decisively rule out a low ETC scale. We will see that this
is far from the case; although the allowed parameter region is
constrained, it could easily be absorbed into strong interaction

unknowns.
We conclude that a low technicolor scale is not very problematic,
and is in fact difficult to probe with the precision necessary
to decisively study extended technicolor interactions of the top.

The plan of this paper is as follows.

We first
consider the effect of ETC interactions that give the top
quark a mass on the $\rho$ parameter. We then review the source
of flavor changing neutral currents in ETC theories.

We  study the consequences
for $B$ mesons. We find that there are big flavor changing effects
in this sector; unfortunately, most will probably not be measured

in the immediate future. The contribution of ETC
interactions to  $B$ meson
mixing is  however quite large.   If there
are indeed additional ETC generated interactions,
constraints on KM angles would be

very different.   We  also study flavor changing effects
in the kaon system. These are consistent with current observations,

although the favored parameter region from

the $\epsilon$--parameter constraint differs from the
standard model.

\section{Rho Parameter Constraint}

 ETC interactions   generate the  top quark mass
 through  an operator of the form

\beq  {g^2_{ETC}

\over M^2_{ETC}} (\overline{\psi^i_L} \gamma^\mu  T_L^{iw})
(\overline{U}^w_R \gamma_{\mu} t_R)+\ {\rm h. c.}
\eeq
where $\psi$ and $t$ are quark fields and $T$ and $U$ are technicolor

fields.
Assuming a single doublet of technifermions, and a custodial SU(2)

broken
only by the flavor dynamics, one can conclude  after a Fierz
transformation that
\beq
m_t \approx {g^2_{ETC} \over M^2_{ETC}} \left(4 \pi v^3\right)
\eeq

Now let us consider an operator generated by ETC exchange between

right
handed $U$ quarks, namely
\beq
{g^2_{ETC} \over M^2_{ETC}} (\overline{U}_R \gamma_{\mu} U_R)

(\overline{U}_R \gamma^\mu U_R)
\eeq
As demonstrated in a different context in ref. \cite{ctsm1},
this gives the
 $Z$ but not the $W$ a mass. This is similar to
the constraint on CTSM models discussed in ref. \cite{ctsm1}.
This was also previously considered in ref. \cite{appel}.
Of course, there is
no way to evaluate this four fermion operator. If for the purpose of
approximation, we factorize the operator into the product of two

right handed currents, we get
\beq
\Delta M_Z^2={g_{ETC}^2 \over M_{ETC}^2} {e^2 v^2 \over 8 \sin^2

\theta_W \cos^2 \theta_W}
\eeq
which, using the equation above, translates into
\beq
\Delta \rho = {m_t \over 8 \pi v} = 1.6 \% {m_t \over 100 \gev}
\eeq
 From Peskin and Takeuchi the 2$\sigma$ limit on $\Delta \rho$

\cite{pt2}
from the $W$ mass alone (assuming zero $S$) is 1\%.
Of course, given strong interaction uncertainties, one cannot
make firm conclusions, but this is as good as any current bound
on the minimal technicolor model with a top quark mass generated by

ETC
interactions. It is clearly a stronger constraint than the limit

generated
by the $Z$ coupling to $b$ quarks, although this would supplement the
current constraint.

In this section, we considered a particular constraint on technicolor
models from precisely measuring electroweak parameters. In the
remainder of this paper, we consider constraints imposed from

studying flavor changing neutral currents.

\section{Flavor Changing Neutral Currents from ETC}
It is easy to see why extended technicolor generates flavor changing

effects.
If there are different extended technicolor scales for the different
quarks, there is no longer a GIM \cite{gim} cancellation mechanism.

Flavor changing
effects are not cancelled.

There are several types of nasty vertices generated by exchange of

ETC gauge bosons.

The first   generated by exchange of ETC gauge bosons
between light quarks produces a vertex of the form
\beq
{g_{ETC}^2\over M_{ETC}^2} (\overline{\psi} \gamma^\mu \psi)^2
\eeq
where $g_{ETC}$ is the coupling of the ETC gauge boson and $M_{ETC}$

is its mass.
Because in general the  states appearing in the above interaction
are not  the mass eigenstates, one can expect flavor changing
effects of size $g_{ETC}^2/M_{ETC}^2$ times the appropriate mixing

angles, depending
on the identity of $\psi$ and the particular flavor changing current

under
consideration.  How big is this effect? If we ignore factors of order

unity,
the constraint that $m_{\rm quark}\approx (g_{ETC}^2/4 \pi v^3

M_{ETC}^2)$ (assuming a minimal technicolor model)
implies flavor
changing neutral current four fermion operators
 of size $m_{quark}/4 \pi v^3$ times appropriate mixing angles, which
 we assume to be approximately the same as the associated KM mixing

angle.

 In addition, there will be direct flavor changing couplings of the
$Z$.
 These arise because the operator in which a light fermion and

technifermion
 current are multiplied translates into a coupling of the $Z$ to the
 light fermion current. Proceeding as in refs. \cite{mesek,lizsek},

one
 finds a $Z$ coupling of size
 \beq
 {g_{ETC}^2 v^2 \over 4 M_{ETC}^2} \overline{\psi_L}
 \left({e \over \sin \theta \cos \theta} Z\!\! \!\!\slash \tau^3
 \right)\psi_L = \xi {m_t \over 16 \pi v} \overline{\psi_L}
 \left({e \over \sin \theta \cos \theta} Z\!\! \!\!\slash \tau^3
 \right)\psi_L.
 \eeq
 Here $\xi$ parameterizes the uncertainty due to
 strong interactions and model dependence.

 Once again, because it is not the mass eigenstate
 which couples in this operator, one expects flavor changing effects
of order $m_t/16 \pi v$ times mixing angles.  We will focus on
flavor changing effects in the down quark sector.

 Here we are assuming that mixing effects do not cancel,
 and are of the order of the associated KM angles.
  There may be equally large effects

associated with light quark flavor physics
 which we do not consider;
 as stated our purpose is to estimate the minimum expected flavor

changing
 effects.

  Flavor changing neutral currents in the standard model generally

arise at one loop.
   The GIM mechanism necessitates the presence
 of a logarithmic or power dependence on the quark mass. Neutral

flavor changing
 currents involving $b$ quarks generally have the largest

contribution
 from a one--loop diagram with an intermediate top quark.  The size

of
 these effects is generally suppressed by $1/16 \pi^2$ times some

function
 of $m_t/v$ which is of order unity for top quark mass between 100

and 200 GeV.
 Some flavor changing effects in the kaon system receive their

dominant
 contribution through diagrams with an intermediate $c$ quark, while
 in general CP violating and GIM doubly suppressed contributions

receive
 the dominant contribution from diagrams with an intermediate top

quark.
 Included in this category is CP violating neutral kaon mixing, CP

violating
 neutral kaon decay, and $K \to \pi \nu \overline{\nu}$.

 For those processes which are dominated by diagrams with an

intermediate
 top quark in the standard model, we expect the ETC generated flavor
 changing effect to be bigger than
  or at least comparable to the standard model
 contribution. This is because of the factor of $1/4 \pi$ relative to

$1/16 \pi^2$. (The dependence on $m_t/v$ is irrelevant to an

estimate
 since this number is of order unity.) We therefore expect large
 flavor changing effects, even with our very conservative assumption
 that we only look at those

 effects generated by mixing with the third generation.

 Notice the original analysis of flavor changing neutral currents
 looked at the effects from the mass generation of light flavors.

Flavor
 changing effects from ETC which are linear in the mass were much

greater
 than those from the standard model which were quadratic in mass.

 For the top quark however, the power dependence
 on the mass is not so dramatic since the mass of the top quark is
  close to $v=246 {\rm GeV}$. The enhanced flavor changing effects
 is due  to
  the factor of $1/4 \pi$ compared to the standard model loop factor

of $1/16 \pi^2$. For this reason, we expect the contribution from ETC

interactions to dominate the standard model loop contribution
 to flavor changing effects in many cases. However, to ultimately
 constrain technicolor models would require effects an order
 of magnitude bigger than those in the standard model, due
 to the uncertainty in our estimates. We will find that
 although ETC effects are large, they are not sufficiently
 large to conclusively constrain ETC models.

 \section{Flavor Changing Neutral Currents and $B$ Mesons}

 We first analyze the flavor changing effects in the $B$ meson

system.
 We consider    flavor changing effects in
 $Z$ decays and in   $B$ meson decays.
 We show in

this section that they  could be  larger
 than in the standard model;

the deviations from the standard model might however

prove difficult to measure.

 First we consider flavor changing $Z$ decay.
 The ratio of the flavor changing partial width between a $b$ and $s$

to the standard model $b$ partial width  is
 \beq
{\Gamma_{Z \to \overline{b} s} \over \Gamma_{Z \to \overline{ b} b}}=
\xi^2 \left(m_t \over 16 \pi v \right)^2 s^2_{23}

\left[ \left({1 \over 2 } -{1 \over 3} \sin^2 \theta_W\right)^2+
\left(\sin^2 \theta_W \over 3
\right)^2 \right]^{-1}
\eeq
 where $s_{23}$, describing the mixing
 between the second and third generations,
 is probably of order $V_{ts}$, and $\xi$ is a constant of order

unity.
 So given the branching ratio into $b \overline{b}$ of 378 MeV,
 we conclude that $Z$ will decay into $b$ and $s$ with branching

fraction

$  10^{-6}(m_t/100\gev)^2$
 where we have taken  $V_{23}=.05$.

 Although  for large top quark mass, this
 could yield some events
 at LEP, certainly more than the standard model,
 it will probably be too difficult to do the flavor identification

necessary
 to ascertain that the flavor changing decay has occurred.

With very
optimistic assumptions, Cocolicchio and Dittmar \cite{cd} conclude
that a branching ratio of $3 \cdot 10^{-5}$ could be observable,
and a possible upper bound limit of the branching ratio of $10^{5}$
might be possible with $10^8$ $Z$'s.  If $\xi$ is large, the flavor
changing decay might be just barely observable, but it is unlikely.

We now
consider flavor changing decays of the $b$.

If we approximate $V_{bc} \approx V_{23}$
 we find that the inclusive rate for hadronic flavor changing decays
 is approximately
 \beq
 \xi^2 10^{-4} \left({m_t \over 100} \right)^2 {\cal P}
 \eeq
 where ${\cal P}$ is a phase space factor to take into account the
 difference between states containing $c$ and $s$ quarks and the

other
 final state particles, and again the factor $\xi^2$ should be
of order unity.

 This is a substantial fraction of flavor changing decays; however
 the branching ratio to exclusive modes is significantly reduced.
 Moreover, if we consider a decay like $b \to s q \overline{q}$,
 it can get a significant contribution from a strong penguin diagram

\cite{gilmanref}.
 Although there is a loop suppression factor, one can expect a

logarithmic
 enhancement of $\ln(m_t/m_c)^2$. This standard model contribution
 would then dominate the flavor changing effect given above.

 We can also estimate the  contribution to

the decays of the form $B \to X_s e^+ e^-$.
 We employ the analysis of Grinstein, Savage, and Wise

\cite{grinwise},
 who computed  the quark decay, and took the ratio of this decay to

the decay to a    final
 state of  the form $X_c e^+ e^-$.

 This yields,  assuming the same mixing angles for both decays,

\beq
 {1 \over (B \to X_c \overline{e \nu}_e)} {d \Gamma

(B \to X_s e^+ e^-)\over d \hat{s}}=
{4 \cdot 10^{-4}\xi^2 \over f(m_c/m_b)} (1-\hat{s})^2 (1+2

\hat{s}^2)

\left({m_t \over v} \right)^2
 \eeq
 where $f(x)=1-8x^2+8x^6-x^8-24x^2 \ln x$ gives the phase space

suppression
 of the $c$ quark and $\hat{s}$ is the $e^+$ $e^-$
 invariant mass divided by the
 square of the $B$ mass. If, as in \cite{grinwise}, we
 restrict the $e^+ e^-$ invariant mass to
 be greater than
  $.2 m_b^2$ to avoid background,
 we get a branching ratio of $3\cdot 10^{-6}(m_t/100 \gev)^2$.

 The standard model
 branching ratio lies between $5\cdot 10^{-7}$ and $3\cdot 10^{-6}$

for top quark
 mass between 50 and 150 GeV. Even if this decay is detected, it

might
 be difficult to distinguish the ETC rate from that
 in the standard model until the top quark mass is known.
 However, if the dependence on $\hat{s}$ could be measured, the two
 contributions would be readily distinguished, since in the
 standard model, the distribution peaks at low invariant mass of
 the $e^+ e^-$. The rate of exclusive modes will probably only
 be a few per cent of the above rate, so it could be difficult
 to measure, depending on the number of $b$'s studied in the future.

 So we see that there is a large contribution to $b$ flavor changing
 neutral currents,

but they are generally comparable
 to the standard model contribution.
 It would therefore be difficult to determine whether

flavor changing effects arise from nonstandard model physics.
 With knowledge of the top quark mass, the situation would however
 be improved.

 We now investigate the constraint in the $B$ meson
 system coming from the more conventional
 tree exchange of an ETC gauge boson between light fermions. As is
 well known \cite{wellknown}, this generates flavor changing

currents.
 This is analagous to the powerful flavor changing neutral current
 constraint in the kaon system, which motivated the more complicated
 technicolor scenarios. Here, as explained earlier, we only look
 at the constraint from the ETC interactions which give rise to
 the top quark mass. We look at the constraint from $B-\overline{B}$

mixing.
 We find a very limited parameter
 space which gives the correct amount of mixing, given current
 bounds on KM angles.
  The conventional standard model constraint on the top quark mass
 and KM angles is \cite{nir}
 \beq
 x_d=\tau_b {G_F^2 \over 6 \pi^2} \eta m_B (B f_B^2) M_W^2 (-E(x_t))
 |V_{td}^* V_{tb}|^2
 \eeq
 where $x_d$ is the $B$ meson mixing parameter, which we take to lie
 between

$0.53$  and $0.70$ \cite{krohu}
  $\eta$
 is the strong interaction correction factor which is approximately

0.85
\cite{nir},
 $B f_B^2$ parameterizes the ratio of the matrix element of the

$\Delta b=2$
 operator to its vacuum insertion value and we take to lie between

0.01 and
 0.04 $\gev^2$, and $V_{td}$ is the KM matrix element.

 The contribution to $B-\overline{B}$ mixing from ETC interactions

will yield
 \beq
 x_d = \tau_b  B f_B^2 m_B \chi {m_t \over 6 \pi v^3}
  |V_{td}^* V_{tb}|^2
 \eeq
 where $\chi$ is a parameter incorporating all uncertainties in this
 equation, including the fact that we take matrix elements of
 different handedness operators to be the same,
  the angles are not really the KM angles, there can be
 group theory factors in the four fermion operator, and the
relation between ETC parameters and the top quark mass is subject
to strong interaction uncertainties and model dependence.
 Comparing the contributions from the ETC vertex and from the

standard model,
 one finds the ETC contribution is about 15--20 times the standard

model
 contribution for all values of the top quark mass between 100 and

200 GeV
 when we take $\chi=1$.
 We therefore neglect the standard model contribution.

 We use the approximate Wolfenstein parameterization of the KM matrix
 \cite{wolf},
 given below.
    \[
  \left( \begin{array}{ccc}
  1-{\lambda^2 \over 2}& \lambda & A \lambda^3 (\rho - i \eta + i

\eta

{\lambda^2 \over 2}) \\
  -\lambda & 1-{\lambda^2 \over 2}-i \eta A^2 \lambda^4 & A \lambda^2

(1+i \eta \lambda^2) \\
  A \lambda^3(1-\rho-i \eta) & -A \lambda^2 & 1
  \end{array}
  \right) \]
 The parameter $\lambda$ is 0.22.
  We take
 $V_{cb}$ between 0.44 and 0.54 \cite{zaitsev}
 and
 $\tau_B=1.4 \pm 0.45 \ {\rm psec} \cite{lep}$. This gives $0.91 < A

< 1.12$.

 Putting all this together yields
 \beq
 0.03 < ((1-\rho)^2+\eta^2) {m_t \chi \over 100 \gev} <0.24
 \eeq
 However, we also know from direct study of the $b$ to $u$ matrix

element
 that $.07 < {|V_{bu}|\over| V_{bc}|}< .13$ \cite{pdb},

which translates into
 \beq \label{above}
 0.10 <\rho^2 +\eta^2 <0.35.
\eeq

 The two above inequalities are just barely
 compatible, since according to eqn.  \ref{above},  the minimum

value of $(1-\rho)^2 +\eta^2$ is 0.17.
This is shown graphically in Figure 1, where we impose the constraint
from $B$ meson mixing simultaneously with the constraint from the

$b$ to $u$ mixing angle.

The dash dot line is the $b$ to $u$ constraint,
while the dashed line is the mixing constraint. We see that
there is only a very small region of overlap.  If the top mass
were bigger than 100 GeV, the overlap would disappear.  Of course,
this is under the assumption that the uncertainty factor is
unity.  The radius of the  semicircles

determined by the $B$--$\overline{B}$ mixing constraint expands
if the factor $\chi$ is small.  Of course, if it is sufficiently
small, the standard model contribution dominates. However we
expect this not to be the case.
Although the exact numbers will change
as the measurements of $b$ parameters improve,
one would expect that the qualitative result of the much stronger

constraint
on KM angles from $B$ meson mixing
will survive because of the large coefficient of the four fermion

operator.
In the standard model, a much larger region of parameter space is

allowed.
In the next section, we put the constraint from neutral $B$ meson
 mixing together with contraints and predictions in the kaon sector.

 \section{Kaon Flavor Changing Neutral Currents}

 We first investigate the constaint from $\epsilon$, the parameter
 of CP violating mixing in the neutral kaon system. Ordinarily,
 the dominant contribution comes from a box diagram with internal
 top quarks, with a nonnegligible contribution from a diagram
 with intermediate charm quarks.  Here, as in the case for mixing

in the $B$ meson system, the nonstandard contribution will dominate.
We again approximate the angles by the KM angles, to get the

constraint
 \beq
 |\epsilon|={(8/3 f_K^2 m_K B_K )\over \sqrt{2} \Delta m_K}
 \left ({m_t \chi' \over 16 \pi v^3} \right) {\rm Im}
  \left(V_{td}^* V_{ts} \right)^2.
  \eeq
  We allow the parameter $B_K$ to vary between $0.3$ and $1.0$.
  There is a further uncertainty in that we have taken the matrix
  elements of operators with different handedness to be the same,
  which is absorbed into the factor $\chi'$ along with the
  other unknowns.
  We combine this constraint with the constraint from $B$--

$\overline{B}$ mixing
  and $V_{bu}$ in Figure 1.

  We observe that for $\chi'=\chi=1$,
  there is only a small region of overlap
where all constraints are simultaneously satisfied. In fact
  for top quark mass only a little greater than 100 GeV, the region
  of overlap would disappear entirely. Of course, due to the various

uncertainty
  factors, this cannot be taken as hard evidence that this additional
  operator is excluded. We do see however that a small value
of the Wolfenstein parameter
$\eta$ is favored, due to both mixing constraints.
We also
  see that the theory with this additional operator is at the

borderline
  of being acceptable.

  We conclude this section by briefly considering two other
processes for which one might have expected the existence of the
low ETC scale for top to be important.
  We are interested in those processes whose primary contribution
  comes from a diagram with an intermediate top quark.

  The potentially most interesting such processes (in terms of

testing
  for this new operator) would be $K \to \pi \nu \overline{\nu}$
  and $\epsilon'$.  In both cases, there are two important effects
  from the additional ETC operator. The first is the much more
  stringent constraint on KM angles, which suppresses both processes.
  The second is the direct additional contribution to flavor changing

$Z$
  vertex. This enhances $K \to \pi \nu \overline{\nu}$, but probably
  not enough to distinguish the model from the standard model
  contribution given the counter effect of the reduction of angles.
  Because the operator which contributes to CP violating kaon decay
  is of the form of the operator produced from electroweak penguin

diagrams,
  it is expected it will work in collaboration with the suppressed
  angles to suppress $\epsilon'$. We briefly elaborate on these

points.

  The operator that is generated by the ETC interaction is
  \beq
  4 \overline{s_L} \gamma_\mu    d_L \overline{\nu_L} \gamma^\mu

\nu_L
  \eeq
  where in our convention these are left handed fermion fields.
  The coefficient of this operator coming from ETC interactions will

be
  \beq
  \xi {m_t \over 32 \pi v^3}

\eeq
  multiplied by mixing angles to the first and second generations.
  Even for a top quark mass of 100 GeV, this generates a coefficient

for
  the relevant operator (with the angles factored out) which is twice

as
  big as the maximum standard model top quark contribution, and

bigger
  still than the $c$ quark contribution. The branching ratio

with the standard model contribution neglected is
  \beq
  B(K^{\pm} \to \pi^{\pm}\nu \overline{\nu})= 6.2\cdot
	    10^{-6}  {|V_{ts} V_{td}^*|^2 \over |V_{us}V_{ud}^* |^2}
  \eeq
  which is less than $2 \cdot 10^{-11}$

  if we incorporate the previous constraint
  on KM angles. So although the coefficient is  large, the rate
  is too small to distinguish this model from the standard model,
  where the prediction for the branching ratio varies between
  $1.5\cdot 10^{-11}$ and $2.5\cdot 10^{-10}$.

  Finally, we consider the prediction for $\epsilon'$. In general,
  $\epsilon'$ is a poor distinguisher of nonstandard model physics
  since it not well determined in the standard model due
 to the uncertainty in matrix elements. However,
 it is very likely that if the KM matrix is the only
 source of CP violation that $\epsilon'$ will be small,
 because
  of the small value of the Wolfenstein parameter, $\eta$.

  Numerically,
  \beq
  {\epsilon' \over \epsilon}=2.3 \cdot

10^{-3} \left( {\tau \over 2\cdot 10^{-4}} \right)
  \left({ \tilde{c}_6 \over 0.1} \right) B_6 \left({175 \mev \over

m_s}
\right)^2
  \left(1-\Omega \right)
  \eeq
  where $\tau=A^2 \lambda^5 \eta$,
$\tilde{c}_6$ is the coefficient of the strong penguin
  operator with the angles factored out,

$B_6$ , which
should be of order unity, is the ratio of the true strong penguin

matrix
  element to the vacuum insertion matrix element, and $\Omega$ is the

fraction
  of the contribution to $\epsilon' / \epsilon$ not coming from
  the strong penguin.
  If $\eta < 0.1$, then $\tau< 5.0 \cdot 10^{-5}$. The
  value of $\tilde{c}_6$ depends on $\Lambda_{QCD}$ but is certainly
  less than 0.1

at a renormalization scale, $\mu$, of $1 \gev$.
The contribution to $\Omega$ from the electromagnetic
  penguin operator, with the additional contribution from ETC

interactions,
  is very large; it is comparable in size to the strong penguin

contribution.
  To be specific, following the analysis of ref. \cite{us},
$\Omega_{EMP}=-0.89[-1.9]$ for $m_t=100[200] \gev$,
  assuming the angles in the down sector are as in the up sector,
the parameter $\xi$ is 1, (four flavor)$\Lambda_{QCD}=100 {\rm MeV}$,
and $\mu=1\gev$. If however, the parameter is $-1$ (or
  equivalently, the product of angles takes the opposite
  sign), one
  finds $\Omega_{EMP}=1.3[2.3]$, again for top quark mass of

$100[200] \gev$
  where we have taken

the ETC induced contribution

to $c7$, the coefficient of the electromagnetic penguin

operator, normalized as in ref. \cite{us} to be

$c_7=\xi {m_t \sin^2 \theta_W /( 12 \pi v)}$. Here we have
  included only the ETC contribution. The standard model piece

which was previously neglected will
  decrease $\Omega_{EMP}$; for example, with $\xi=-1$ and the angles
  as for KM angles, $\Omega_{EMP}=1.6$ for $m_t=200 \gev$. It is

unlikely that $\epsilon'/\epsilon$
  will exceed $5 \cdot 10^{-4}$.

  \section{Conclusion}
  We have considered the effects of ETC interactions which give
  rise to a top quark mass. We have found that the deviation in the

$\rho$
  parameter from unity is quite large.
  However, it remains possible that the top ETC scale can be
  as low as would be required to generate a top quark mass
  greater than 100 GeV.
  If large deviations of the $Z b \overline{b}$ vertex or
  the $\rho$ parameter are observed, it could prove to
  be positive evidence of nonstandard physics, but
  it seems difficult to rule out a low ETC scale from not
  observing devations.

  We have found furthermore that flavor changing constraints are

stringent,
  but again do not conclusively rule out ETC models.
  Here, the strongest constraint comes from simultaneously
  satisfying the $B-\overline{B}$ mixing and $b$ to $u$
   constraints.

  A large value of $\epsilon'$ might
  be a good indication that there is no low top ETC scale;
  however it might also just be evidence of a source other
  than the KM matrix of CP violation.

  So it is clear that a low top scale which violates flavor
  symmetry is allowed, and
  furthermore, that it will be difficult to test. However, in
nonminimal
  models, the contribution might be larger than we estimate

\cite{cdg}.

The best constraint, since it is independent of mixing
angles, is probably the $\rho$ parameter constraint.  In models
with a common ETC scale for the up and down sector

\cite{randall}, even this constraint is quite weak.

The situation for light quarks is very different. Here it is clear

that

ordinary ETC does not work because ETC induced flavor changing
effects far exceed the standard model loop induced effects.

Some mechanism must
  be present to turn linear dependence on the quark mass into
  quadratic dependence; whether it is a GIM like mechanism or
  a walking scenario remains to be seen. In ref. \cite{randall},
  models are constructed which respect a

GIM mechanism. Even for these models, the flavor violating
   effects which are
  considered in this paper will be relevant because of the large
  flavor symmetry breaking which is required in order to generate a

top
  quark mass.

  \section*{Acknowledgements}
  I am grateful to Hitoshi Yamamoto and Bolek Wyslouch for useful

information.
  I thank Elizabeth Simmons for comments on the manuscript.
  This work is supported in part by funds provided by the U. S.

Department
  of Energy (D.O.E.) under contract \#DE--AC02--76ERO3069.

\def\thefiglist#1{\section*{Figure Captions\markboth
 {FIGURE CAPTIONS}{FIGURE CAPTIONS}}\list
 {Figure \arabic{enumi}.}
 {\settowidth\labelwidth{Figure #1.}\leftmargin\labelwidth
 \advance\leftmargin\labelsep
 \usecounter{enumi}\parsep 0pt \itemsep 0pt plus2pt}
 \def\newblock{\hskip .11em plus .33em minus -.07em}
 \sloppy}
\let\endthefiglist=\endlist

\begin{thefiglist}{9}
\item Constraints on KM angles with a low ETC scale to generate the
top quark mass (here
taken as 100 GeV)
 included
 with strong uncertainty factors taken to be 1.
 The solid line is the constraint from $\epsilon$,
the dot--dash line is the constraint from the $b$ to $u$ mixing

angle,
and the dashed line is the constraint from $B$--$\overline{B}$

mixing.
\end{thefiglist}

 \end{document}